\documentclass[aps,prl,twocolumn,showpacs]{revtex4}
\usepackage{epsfig}
\usepackage{amsbsy,latexsym}
\usepackage{amsmath}
\usepackage{amssymb, mathrsfs}
\usepackage[mathscr]{eucal}

\def\d{\operatorname{d}}\def\<{\langle}\def\>{\rangle}
\def\Tr{\operatorname{Tr}}\def\:{\hbox{\bf :}}
\def\conv#1{\mathscr{#1}}\def\vec#1{{\boldsymbol{#1}}}

\def\geq{\geqslant}\def\leq{\leqslant}
\def\>{\rangle}
\def\<{\langle}
 
\def\sH{\mathcal{H}}\def\cP{\conv{P}}
\def\dim{\operatorname{dim}}\def\Klm{\Omega}

\def\qed{$\,\blacksquare$\par}

\newtheorem{lemma}{Lemma}

\newtheorem{theorem}{Theorem}
\def\Proof{\medskip\par\noindent{\bf Proof. }}

\begin{document}

\title{How continuous quantum measurements in finite dimension are actually discrete} 
\author{Giulio Chiribella}\email{chiribella@fisicavolta.unipv.it} 
\affiliation{{\em QUIT Group}, Dipartimento di Fisica  ``A. Volta'' and INFM, via Bassi 6, 27100 Pavia, Italy}
\homepage{http://www.qubit.it}
\author{Giacomo Mauro D'Ariano}\email{dariano@unipv.it}
\affiliation{{\em QUIT Group}, Dipartimento di Fisica  ``A. Volta'' and INFM, via Bassi 6, 27100 Pavia, Italy}
\homepage{http://www.qubit.it}
\author{Dirk Schlingemann}
\email{d.schlingemann@tu-bs.de}
\affiliation{Institut f\"ur Mathematische Physik, Technische Universit\"at Braunschweig,
  Mendelssohnstra{\ss}e~3, 38106 Braunschweig, Germany} 
\date{\today}

\begin{abstract}
We show that in finite dimension a quantum measurement with continuous
set of outcomes is always equivalent to a continuous random choice of
measurements with only finite outcomes.
\end{abstract}
 \maketitle When we measure the spin component along a magnetic field with
a Stern-Gerlach apparatus, for spin
1/2 particles we have only two possible outcomes: spin up and spin down. This measurement is perfectly
repeatable, and can perfectly discriminate between the two orthogonal
states $|\!\!\!\uparrow\>$ and $|\!\!\!\downarrow\>$.  It is possible,
however, to design an experiment with more than two outcomes, which
discriminates optimally---though not perfectly---among three or more
non orthogonal states. Indeed, a four-outcome measurement on
a two-level system is needed in the eavesdropping of a BB84
cryptographic communication\cite{Nielsen2000}, or in the lab to perform  an {\em informationally
  complete measurement}\cite{InfoCompletePOVMs}, which determines the quantum state from the
measurement statistics. 

\par 
What about performing a measurement with a {\em continuous} set of
outcomes? This is the case of a measurement designed to optimally
determine the ``direction'' of a spin\cite{SpinCoherent},
similarly to what we do in classical mechanics. Such a measurement
produces a probability $p({{\vec n}})\d{{\vec n}}$ of the spin
direction falling within the solid angle $\d{{\vec n}}$ around the
direction ${\vec
n}=(\sin\theta\cos\phi,\sin\theta\sin\phi,\cos\theta)$. Indeed the
measurement of direction must be feasible\cite{spindir}---though in-principle
inaccurate---otherwise Quantum Mechanics would fail in describing what
we normally observe in the macroscopic world.  Actually, this is not the
only interesting example of continuous-outcome measurement on a
finite-level system: in fact, measurements of this kind have an endless number of applications, e.g.  optimal state estimation\cite{MassPop}, optimal alignment of
directions\cite{direction} and reference frames\cite{refframe},
optimal phase estimation\cite{Holevo}, and optimal design of atomic clocks\cite{OptClocks}.

\par
In this Letter we establish a fundamental property of quantum
measurements with continuous set of outcomes, namely that for finite
level systems any such measurement is equivalent to a continuous
random choice of measurements with only \emph{finite} outcomes.  This
means that any physical quantity measured on a finite dimensional
system is intrinsically discrete, while the continuum is pure
classical randomness.  For a spin 1/2 particle, this fact is well
illustrated by the simple observation that the optimal measurement of
direction can be equivalently realized by a customary Stern-Gerlach
experiment where the magnetic field is randomly oriented. We emphasize
that, in general, the discretization of physical quantities does not
involve just von Neumann observables, but, more generally, finite
measurements with a number of outcomes larger than the Hilbert space
dimension. Morever, using the main result we show that any continuous
measurement that optimizes some convex figure of merit
(e.g. maximizing the mutual information or the Fisher information, or,
alternatively, minimizing a Bayes cost\cite{Helstrom,Holevo}) can be
always replaced by a
\emph{single}
measurement with finite outcomes, without affecting optimality. 


Let us start by briefly reviewing the general theoretical description
of measurements in Quantum Mechanics.  Consider a quantum system (with
Hilbert space $\sH$ of dimension $\dim(\sH)=d<\infty$), which
undergoes a measurement with random outcome $\omega$, distributed in
the outcome space $\Omega$. The probability distribution of the
outcomes for any possible state $\rho$ of the system depends on the
specific measuring apparatus used, and is given by a {\em positive
operator-valued measure} (acronym POVM), namely the probability that the
outcome falls in the subset $B
\subseteq \Omega$ is given by the Born rule $p(B)=\Tr[\rho P(B)]$, where $P(B)$ is a nonnegative operator with
normalization condition $P(\Omega)=I$\cite{Nota:Povm}.  In the special
case in which the measurement is finite, a random result $i$ from a set of possible outcomes $\{i=1,2,\ldots, N\}$ is returned with
probability $p_i=\Tr[\rho P_i]$, $P_i\geq 0$ being nonnegative operators with normalization condition $\sum_{i=1}^N
P_i=I$.

Before presenting the main result, in order to help intuition, we briefly analyse two simple
prototypes of continuous measurement: the optimal measurement of the "spin direction" for a spin 1/2
particle, and the optimal measurement of a phase shift.

\par The measurement of direction for spin 1/2 particles is given by the POVM \cite{SpinCoherent}
\begin{equation}\label{spindirPOVM}
P(B)=\int_B \frac{\d{\vec n}}{2\pi}|{\vec n}\>\<{\vec n}|,
\end{equation}
where $|{\vec n}\rangle$ is the eigenvector of 
$\vec n \cdot {\vec J}$ with eigenvalue $+1/2$, $\vec J$ being the spin operator.  
It is simple to see that this measurement is equivalent to the randomization  
\begin{equation}\label{SpinRand}
P(B)=\int_{\mathbb{S}^2}\frac{\d{\vec n}}{4\pi}\,E^{({\vec n})}(B)~,
\end{equation}
where $\d \vec n /(4\pi)$ is the uniform probability distribution over the unit sphere $\mathbb{S}^2$ and  $E^{(\vec n)}$ is the POVM 
\begin{equation}
E^{({\vec n})}(B)=\chi_B({\vec n})|{\vec n}\>\<{\vec n}|+\chi_B(-{\vec n})|-{\vec
  n}\>\<-{\vec n}|~.\label{extrspin1}
\end{equation}
[$\chi_B(\vec n)$ is the characteristic function of the set $B$:
 $\chi_B(\vec n)=1$ for $\vec n\in B$, $\chi_B({\vec n})=0$
otherwise.]  The POVM $E^{(\vec n)}$ represents a measurement of
direction based on a Stern-Gerlach setup with magnetic field oriented
along $\vec n$: if the apparatus outputs ``up'', one assigns to the spin the
direction $\vec n$, if `down'', one assigns $-\vec n$. With this data-processing, the
probability of observing the spin within the region $B$ is nonzero
only if $B$ contains at least one of the directions $\pm
\vec n$. The continuous POVM (\ref{spindirPOVM}) is then interpreted as a Stern-Gerlach
measurement performed with a direction ${\vec n}$ of the magnetic
field randomly chosen in the unit sphere.  

Another example of continuous measurement is that of phase estimation, where one wants to measure
the phase shift $\phi\in [0,2\pi)$ experienced by a quantum state under the action of the unitary
evolution $U_\phi=\exp(iN\phi)$, with $N=\sum_{n=0}^{d-1}|n\>\<n|$, $\{|n\>\}$ orthonormal basis
for $\sH$. The optimal POVM is given by \cite{Holevo}
\begin{equation}
P(B)=\int_B \frac{\d\phi}{2\pi}|\phi\>\<\phi|,\qquad |\phi\>=\sum_{n=0}^{d-1} ~e^{i n \phi}|n\>.
\end{equation}
and is equivalent to the randomization: 
$P(B)=\int_0^{2\pi} \frac{\d\phi}{2\pi}\, E^{(\phi )}(B)$,
where $E^{(\phi)}$ is the POVM $E^{(\phi)}(B)= \frac{1}{d}\sum_{n=0}^{d-1}\chi_B(\phi_n + \phi)~
|\phi_n +\phi \>\<\phi_n + \phi|~, \phi_n=\frac{2\pi n}{d}$.  

We will now show that all continuous measurements in finite dimension
can be always interpreted in an analogous way, namely as a continuous
random choice of measurements with finite number of outcomes. More
precisely, we will prove the following
\begin{theorem}\label{maintheo}
For any POVM $P(B)$ the following
decomposition holds
\begin{equation}\label{decompos}
P(B)=\int_{\mathcal{X}}\d x\, p(x)~ E^{(x)}(B),
\end{equation}
where $x\in{\mathcal X}$ is a suitable random variable, $p(x)$ a probability density, and, for every value of
$x$, $E^{(x)}$ denotes a POVM with finite support, i.e. of the form 
\begin{equation}
\label{decomposE}
E(B)=\sum_{i=1}^{d^2}\chi_B(\omega_i) P_i
\end{equation}
$\{\omega_i \in \Omega\}$ being a set of points, and $\{P_i\}$ being a
finite POVM with at most $d^2$ outcomes\cite{Nota:NumberOfOutcomes}.
\end{theorem}

A POVM $E$ of the form of Eq. \eqref{decomposE} is in turn nothing but
the continuous data-processing of the finite POVM $P_i$, with function
of the outcomes $f(i)= \omega_i$: if the apparatus outputs $i$, then
one assigns to the measurement the outcome $\omega_i$. The
decomposition (\ref{decompos}) shows that the continuous POVM $P$ is
achieved by randomly choosing a classical parameter $x\in \mathcal{X}$
and then performing the finitely-supported POVM $E^{(x)}$, depending
on $x$ through the finite POVM $\{P_i^{(x)}\}$ and through the points
$\{\omega_i^{(x)}\}$.  Operationally, this corresponds to the
following recipe: {\em i)} randomly draw a value of $x$ according to
$p(x)$; {\em ii)} depending on $x$, measure the finite POVM $P^{(x)}$,
thus getting the outcome $i$; {\em iii)} for outcome $i$, assign to
the continuous measurement the outcome $\omega_i^{(x)}$.  As a first
consequence, this simple recipe shows that, contrarily to a rather
common belief (see, e.g. \cite{DBE}), continuous quantum measurements in finite dimension are as feasible as the discrete ones.

\begin{figure}[h]
\epsfig{file=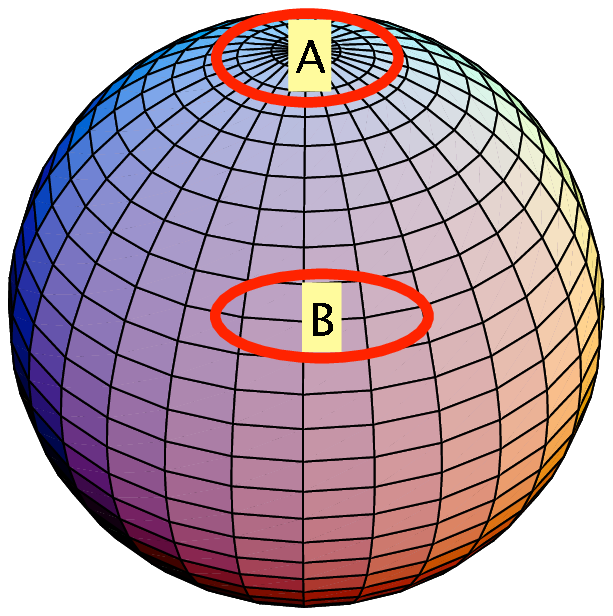,width=4cm}
\epsfig{file=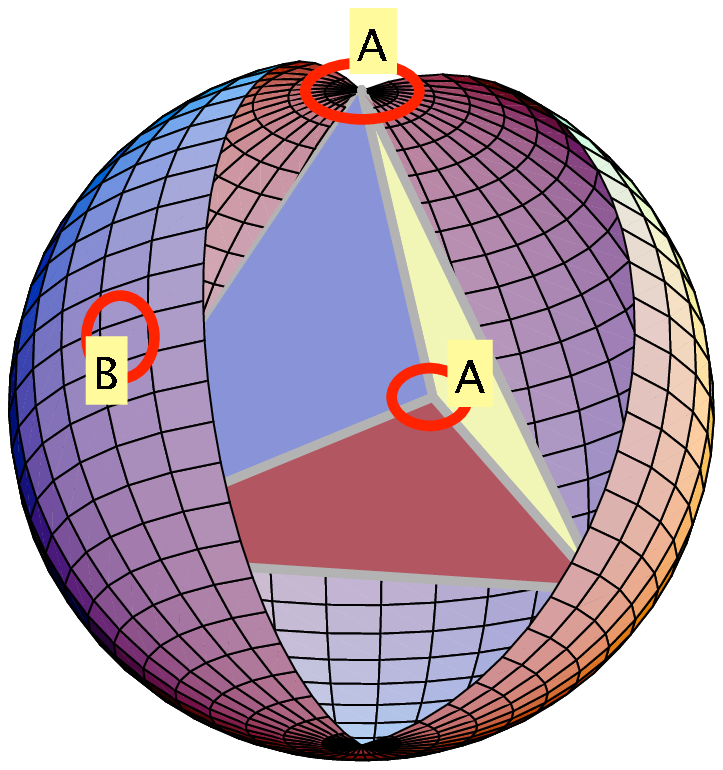,width=4cm}
\caption{{\bf Left:} Illustration of the POVM $E^{({\vec n})}$ in  Eq. (\ref{extrspin1}) as an example of finitely-supported POVM $E^{(x)}$ in Eq. (\ref{decomposE}).
In this specific example the outcome space $\Omega$ is the unit sphere
$\Omega\equiv{\mathbb S}^2$, the dimension of the Hilbert space is $d=2$, and only two out of the four terms in Eq. \eqref{decomposE}
are nonvanishing, namely  $P_1=|\vec k\>\<\vec k|,\, P_2=|-\vec k\>\<-\vec
k|,\,P_3=P_4=0$. The probability of finding the spin direction in a region $R \subseteq \Omega$
 is zero for $R$ missing the two
poles, as  $B$ in the figure, and is possibly nonzero for $R$ as $A$.  {\bf Right:} Another example of finitely-supported POVM for
 $d=2$, corresponding to a SIC (symmetric informationally
complete) POVM\cite{SIC}.  The POVM is made of four elements $P_i$ corresponding
to $\omega_i$ at the vertexes of a tetrahedron. The
probability can be nonvanishing only if the region $R$ contains at
least one of these four points, such as in $A$, whereas it is always
zero in situations as in $B$.}\label{extspn}
\end{figure}

The decomposition of the measurement of the "spin direction" given by
Eq. (\ref{SpinRand}) provides a concrete example of decomposition
(\ref{decompos}). In particular, the finitely-supported POVM
$E^{({\vec n})}$ in Eq.  (\ref{extrspin1}) is illustrated in Fig.
\ref{extspn} for $\vec n= \vec k$.  Notice that, in general, there may be different randomization schemes yielding the same continuous POVM: as an example,
Fig. \ref{extspn} illustrates another finitely-supported POVM that
allows one to reproduce the measurement of direction by simply
randomizing the orientation of the Cartesian axes.


We now derive the main result. We fix both the quantum system and the outcome
space $\Omega$, and consider the set $\cP$ of all possible POVMs for
these. This is a convex set, since given any two POVMs $P'$ and $P''$,
their {\em convex combination} $P^{(\lambda)}=\lambda
P'+(1-\lambda)P''$ for $\lambda\in[0,1]$ is still a POVM, namely the
whole segment joining $P'$ and $P''$ is contained in $\cP$. The {\em
extremal points} of the convex set $\cP$ are those POVMs that cannot
be written as convex combination of two different POVMs. Stated
differently, a POVM $P \in\cP$ is not extremal if and only if it is
the midpoint of a segment completely contained in $\cP$, i.e. if and
only if there exist two distinct points $P', P''\in\cP,~ P'\not= P''$
such that $P=
\tfrac{1}{2}(P' + P'')$. This is equivalent to the existence of a
direction $Q \not=0$ and a positive number $\epsilon >0$ such that
$P + t Q \in\cP$ for any $t \in [-\epsilon, \epsilon]$.
The standard name for the direction $Q$ in convex analysis is
\emph{perturbation}. Here  the perturbation $Q$ is a function that
associates to any subset $B\subseteq \Omega$ an operator $Q(B)$, fulfilling the
three requirements: \emph{i)} $Q(B)$ is Hermitian for any subset
$B\subseteq\Klm$; \emph{ii)} $Q (\Klm)=0$; \emph{iii)} $P(B) + t
Q(B)\geq 0$ for any $B\in\Klm$ and for any $t \in [-\epsilon,
\epsilon]$.

If there exists a nonzero perturbation $Q$ for $P$, then $P$ is non extremal: using this criterion,
we now establish that the extremal POVMs must necessarily have finite support, namely they must be of the form of Eq. \eqref{decomposE}.

The proof takes advantage of the following:
\begin{lemma}
  Every POVM $P\in\cP$ admits a {\em density} with unit trace, namely for any POVM $P$ there exists a finite measure
  $\mu (\d\omega)$ over $\Omega$ such that 
\begin{equation}\label{POVMDensity}
P(B) = \int_{B} \mu (\d \omega)~ M(\omega)~,
\end{equation}
with $M(\omega)\geq 0$ and $\Tr[M(\omega)]=1$ $\mu$-almost everywhere.
\end{lemma}
\Proof Consider the finite measure $ \mu (\d \omega)$ defined by $\mu (B) = \Tr[P(B)] ,~\forall B \subseteq \Omega$~.
Since $P(B)\geq 0$, one has $P(B) \leq\Tr[P(B)] ~I= \mu (B) ~I$, namely $P(B)$ is dominated by the
measure $\mu (B)$.  This implies that $P$ admits a density $M(\omega)$ with respect to $\mu (\d
\omega)$. Clearly, the density $M(\omega)$ has to be nonnegative $\mu-$almost everywhere. Moreover, for any $B\subseteq\Klm$ one has $\int_B \mu
(\d\omega)\equiv\mu(B)=\Tr[P(B)]=\int_B\mu(\d\omega) ~\Tr[M(\omega)]$, whence 
$\Tr[M(\omega)]=1$ $\mu$-a.e.\qed

Thanks to this Lemma, we can represent any POVM $P\in\cP$ using its density $M(\omega)$ as in Eq.
\eqref{POVMDensity}.  To prove that an extremal POVM must be of the form \eqref{decomposE} it is
enough to show that for extremal POVMs the measure $\mu (\d \omega)$ must be concentrated on a
finite set of outcomes $\{\omega_1, \dots, \omega_{d^2}\}$, i.e.  $\mu (B) =0$ for any set $B
\subseteq\Omega$ not containing anyone of the points $\omega_i$.  We recall the definition of the
support of a measure $\mu(\d \omega)$ as the set of all points $\omega\in\Omega$ such that $\mu (B)
>0$ for any open set $B$ containing $\omega$.

\begin{lemma}\label{Lemma:Ext}
  Let $P\in\cP$ be a POVM and $\mu(\d\omega)$ the measure defined by
  $\mu (B) = \Tr[P(B)]$.  If $P$ is extremal, then the support of
  $\mu(\d\omega)$ is finite and contains no more than $d^2$ points.
\end{lemma} \Proof Suppose that the support  contains more than $d^2$ points. In this case,
one can take $d^2 +1$ points $\omega_i \in \Omega$ in the support and $d^2 +1$ disjoint open sets $U_i \subset \Omega$, $i =1, \dots, d^2 +1$, such that  $\omega_i \in U_i$ for any $i$\cite{Nota:Hausdorff}. As a consequence, the space $L^{\infty} (\Omega,
\mu)$ of integrable functions $f(\omega)$ that are bounded $\mu-$almost everywhere has dimension at least $d^2
+1$ (indeed, the characteristic functions $\chi_{U_i} (\omega)$ are a
set of $d^2 +1$ bounded and linearly independent functions). Then,
consider the matrix elements $f_{ij}(\omega)= \<i| M(\omega) |j\>$,
where $M(\omega)$ is the POVM density of Eq. (\ref{POVMDensity}), and
$|i\>, |j\>$ are elements of an orthonormal basis for $\sH$. Since the
operators $M(\omega)$ are nonnegative with unit trace a.e., the
functions $f_{ij} (\omega)$ are bounded a.e., namely $f_{ij} \in
L^{\infty} (\Omega, \mu) \quad \forall i,j$. Moreover, since the space
$L^{\infty}(\Omega, \mu)$ has dimension larger than $d^2$, it must
contain at least one function $g(\omega) \not =0$ that is linearly
independent from the set $\{f_{ij}\}$. Using the Gram-Schmidt
orthogonalization procedure, such a function $g$ can be always chosen
to be orthogonal to all $f_{ij}$, namely $
\int_{\Omega} \mu (\d \omega)~ g^* (\omega) f_{ij}(\omega) =0 \quad \forall i,j$. 
Finally, since $f_{ij}^*(\omega) = f_{ji} (\omega)\quad \forall i,j$,
such a $g$ can be also chosen to be real.  Now, we claim that the Hermitian
operators $Q(B)$ defined by
\begin{equation}\label{Q}
Q(B) = \int_B \mu (\d \omega)~ g(\omega)~ M(\omega)
\end{equation} 
provide a perturbation for the POVM $P$. Indeed,  we have
$Q(\Omega) =0$ as the matrix elements $\<i| Q(\Omega)|j\>$ are zero for any $i,j$:
\begin{eqnarray}
\<i|Q(\Omega) |j\> &=& \int_{\Omega} \mu (\d \omega)~  g(\omega) \<i| M(\omega)|j\>\\
& =& \int_{\Omega} \mu (\d \omega) g(\omega) f_{ij}(\omega) =0~.
\end{eqnarray}   Moreover, since $g \in L^{\infty} (\Omega,
\mu)$, there exists a positive number $c$ such that $|g(\omega)|\leq c< \inftyù$
a.e., thus implying that the operators $M(\omega)~[1 + t g(\omega)]$
are a.e. nonnegative for any $t \in [-\epsilon,\epsilon]$, with
$\epsilon = 1/(2c)$. Hence, integrating over any subset $B$, we obtain
that the operators $P(B) + t Q(B)$ are nonnegative, namely $Q$ is a
perturbation. Finally, $Q$ is nonzero, otherwise taking the trace of
Eq. \eqref{Q}, and using that $\Tr[M(\omega)]=1$ a.e. we would get
$0 = \Tr[Q(B)] = \int_{B} \mu(\d \omega) ~g(\omega) \quad \forall B$, thus implying
$g = 0$, which is not possible by definition of $g$.  In conclusion, if the support of $\mu(\d
\omega)$ contains more than $d^2 $ points, then the POVM $P$ has a
nonzero perturbation, whence it is not extremal.  \qed

Lemma \ref{Lemma:Ext} establishes that an extremal POVM has necessarily the form of Eq.
\eqref{decomposE}, namely it can be realized by measuring a finite POVM $P_i$ and declaring
measurement outcome $\omega_i$. Using this fact, we readily obtain the
proof of the main Theorem: \par {\bf Proof of Theorem \ref{maintheo}.} Due to the
standard Krein-Milman theorem of convex analysis, any point of a
compact convex set is a continuous convex combination of points that
are either extremal or limit of extremals.  On the other hand, it is
simple to prove that the set $\cP$ of all POVMs is
compact\cite{Nota:Compact}, and that any limit of extremal POVMs is
still a POVM of the form
\eqref{decomposE}\cite{Nota:Discrete}.  For a detailed mathematical proof see Ref.
\cite{nextCDS}.\qed

We now want to explore some consequences of decomposition (\ref{decompos}) for
optimization of POVM's and for Quantum Tomography. Optimizing a
quantum measurement consists in finding the POVM $P$ that maximizes
the value of a figure of merit $\mathcal F [P]$---e.~g. mutual or
Fisher information, average fidelity, or any Bayes gain. In all these
cases $\mathcal F[P]$ is convex, i.~e. $\mathcal F [\lambda P' +
(1-\lambda) P''] \le \lambda \mathcal F[P'] + (1-\lambda) \mathcal
F[P'']$ for any $\lambda \in [0,1]$. Suppose now that a continuous
POVM $P$ is optimal for $\mathcal F$. Combining convexity of $\mathcal
F$ with Eq.  (\ref{decompos}) one has
\begin{equation}
\mathcal F_{\max} = \mathcal F [P] \le \int_{\mathcal X} \d x p(x)~ \mathcal F[E^{(x)}] \le \mathcal F_{\max}~,
\end{equation} 
which implies $\mathcal F[E^{(x)}] = \mathcal F_{\max}$ for any $x$ except at most a set of zero
measure. This means that all the finite POVMs $E^{(x)}$ are equally optimal: in particular,
\emph{for any optimal continuous measurement there is always an optimal measurement with finite (no
  more than $d^2$) number of outcomes.}  In special situations some explicit algorithms to find
optimal finite measurements are known\cite{DBE,Latorre,IbRoland}. In particular, Ref.\cite{Latorre}
shows that in many cases the minimal number of outcomes is larger than $d=\dim(\sH)$.  Combined with
the above result, this fact definitely proves that the quantum discretization cannot rely solely on
von Neumann measurements.

Regarding Quantum Tomography, using the present analysis we can make mathematically precise the
common intuition that an informationally-complete continuous measurement is equivalent to a
Tomography scan made of a random choice of observables---or more generally POVMs. Indeed one can
estimate the ensemble average of any operator $A$ by using the two data processing $f_A(\omega)$ and
$f_A(\omega_i^{(x)})$ for continuous POVM and tomography, respectively, as follows
\begin{equation}\nonumber
A=\int_\Omega\d\omega\, f_A(\omega)M(\omega)=\int_{\mathcal{X}}\d x\, p(x)
\sum_{i=1}^{d^2}f_A(\omega_i^{(x)})P_i^{(x)}. 
\end{equation}

In conlusion, in this Letter we showed that continuous quantum
measurements can be always realized by performing finite measurements
depending on a random classical parameter. Physical properties, such
as spatial orientation and time, are then intrinsically discrete when
measured on finite level quantum systems.

\par {\em Acknowledgments.---} This work has been founded by Ministero Italiano dell'Universit\`a e
della Ricerca (MIUR) through PRIN 2005. D. S. acknowledges support by Italian CNR-INFM.
G. C. acknowledges  P. Perinotti, M. Keyl, L. Cattaneo, and S. Albeverio for many useful discussions in the preliminary stage of this work.
     

\begin{thebibliography}{99}
\bibitem{Nota:NumberOfOutcomes} A number of
outcomes less than $d^2$ is equivalent to having some zero operators
$P_i$.
\bibitem{Nota:Hausdorff} The existence of the disjoint open sets $U_i \ni \omega_i$ is guaranteed if the outcome space $\Omega$ is a Hausdorff space, which is always the case for the physical outcome spaces.
\bibitem{Nota:Compact} An operational notion of distance between POVMs is given in terms of
  difference between the pertaining expectations $\<f\>_{\rho} =
  \int_{\Omega} f(\omega) \Tr[\rho P(\d \omega)]$ of a function $f$ of
  outcomes. The Banach-Alaoglu theorem shows that the set of POVMs is
  compact with respect to such distance.
\bibitem{Nota:Discrete} For the easy case of $\Omega$ compact (e.g. spin direction or phase), let
  $E^{(n)}(B) = \sum_{i=1}^{d^2}~ \chi_B (\omega^{(n)}_i)~P_i^{(n)}$ be a sequence of extremal
  POVMs with limit $E(B)$. For any $i$ we can find a subsequence
  $\omega_i^{(n_k)}$ converging to a point $\omega_i \in \Omega$. Similarly, we can choose $n_k$
  so that also the finite POVMs $\{P_i^{(n_k)}\}$ converge to a finite POVM $\{P_i\}$.  Hence $E(B)
  = \sum_{i=1}^{d^2} ~\chi_B( \omega_i) P_i$, namely the limit is still of the form \eqref{decomposE}.
\end{thebibliography}
 \end{document}